\documentstyle[12pt]{article}
\renewcommand{\thefootnote}{\fnsymbol{footnote}}

\begin{document}
\begin{titlepage}
\setcounter{page}{1}
\thispagestyle{empty}
\rightline{\vbox{\halign{&#\hfil\cr
&UH-511-1124-2008\cr
&March 2008\cr}}}
\vspace{0.8in}

\begin{center}

{\Large\bf {NEUTRINO FLAVOR GONIOMETRY BY HIGH ENERGY 
ASTROPHYSICAL BEAMS\\}}
\medskip

{\normalsize \large Sandip Pakvasa}
\\ \smallskip
{\it {Department of Physics and Astronomy\\University of Hawaii,
Honolulu, HI 96822}}\\
\end{center}

\begin{abstract}

It is shown how high energy neutrino beams from 
very distant sources can be utilized to learn about many
properties
of neutrinos such as lifetimes, mass hierarchy,
mixing, minuscule pseudo-Dirac mass splittings and other
exotic properties; in addition, the production mechanism of 
neutrinos in astrophysical sources can also be elucidated. 

\end{abstract}

\vskip0.45in



\renewcommand{\thefootnote}{\arabic{footnote}}
\end{titlepage}


\section{Introduction}
We make several  basic assumptions which are reasonable.  The 
first one is that distant neutrino
sources (e.g. AGN's and GRB's) exist; and furthermore with 
detectable fluxes
at high energies (up to and beyond PeV).  The second one is 
that in the not
too far future, very large volume, well instrumented 
detectors of sizes of
order of KM3 and beyond will exist and be operating; and 
furthermore will
have (a) reasonably good energy resolution and (b) good 
angular resolution
($\sim 1^0 $ for muons).  The first is motivated by the fact that
cosmic rays are observed all the way to $10^5PeV$ and gamma rays to
100 TeV and these sources would presumably produce neutrinos as well.
We further assume that a neutrino signal will be seen with a reasonable
event rate.  Finally, we assume that neutrino flavors can be 
distinguished. At the moment we know how to do this for $H_2O$ 
$\hat{c}$ detectors\cite{icecube}; however,  extending such flavor 
identification to
other types of detectors such as ones based on air shower arrays\cite{auger} or
the Askaryan effect\cite{anita} remains a task for future.

\section{Neutrinos from Astrophysical Sources}

If these two assumptions are valid, then there are a 
number of uses these
detectors can be put to\cite{pakvasa}.  Here I  
want to focus on those that enable us
to determine some properties of neutrinos: probe 
neutrino lifetimes 
to $10^4 s/eV$ (an improvement of $10^8$ over current 
bounds), 
pseudo-Dirac mass splittings to a level of $10^{-18} eV^2$ 
(an improvement of a factor of $10^6$ over
current bounds) and in case of very small pseudo-Dirac mass differences 
measure cosmological parameters such as red-shift in neutrinos. There is
the possibility of potentially  measuring  quantities 
such as $|U_{e3}|$ and the phase $\delta$ in the MNSP 
matrix\cite{MNS}.

\section{Astrophysical neutrino flavor content}

In the absence of neutrino oscillations we expect a very 
small $\nu_\tau$
component
in neutrinos from astrophysical sources. From the most 
discussed and  the most likely astrophysical high energy
neutrino sources\cite{learned} we expect nearly equal 
numbers of particles
and anti-particles, half as many $\nu_e's$ as $\nu_\mu's$ 
and virtually no $\nu_\tau's$.  This comes about simply 
because the neutrinos are thought to originate in decays 
of pions (and
kaons) and subsequent decays of muons.  Most astrophysical 
targets are fairly tenuous even compared to the Earth's 
atmosphere, and would allow for full muon decay  
in flight.  (There could be  flavor 
independent fluxes from cosmic defects and exotic objects 
such as evaporating black holes.  Observation of
tau neutrinos from these would have great importance.)  A 
conservative estimate\cite{learned1} shows that the prompt 
$\nu_\tau$ flux is very small and the emitted flux is 
close to the ratio $1:2:0$.  The flux ratio of $\nu_e: 
\nu_\mu: \nu_\tau = 1:2:0$
is certainly valid for those AGN(or GRB) models in which the 
neutrinos are 
produced in beam dumps of photons or protons on matter, in 
which
mostly pion and kaon decay(followed by the decay of muons) 
supply the bulk of 
the neutrino flux. 

This flavor mix of 1:2:0 is only approximate; a more careful estimate shows
the actual result is 1:1.85: $\epsilon$ where  $\epsilon$ is rather small
(less than 0.001)\cite{lipari}.  The precise mix  also depends on the energy spectrum at
injection.  The sources in which the primary process is $\gamma p$ rather than
pp are distinguished by a lack of $\bar{\nu}_e's$ in the initial flavor mix;
this is due to the fact that $\gamma p$ scattering produces dominantly
$\pi^{+}$ which does not have a $\bar{\nu}_e$ amongst its decay products.

There are two other initial flavor mixes possible.  One is the so called damped
muon case when the $\mu's$ lose energy (via interaction with strong magnetic
fields or with matter\cite{rachen}).  The lower energy of the muon makes 
the $\nu_e$ have  much lower energy
than the $\nu_\mu$ from $\pi$ decay,
and hence effectively  the flavor mix becomes  $\nu_e: \nu_\mu: \nu_\tau = 0:1:0$.  
Again it should be emphasized that this is not exact; the  $\nu_e$ content is 
never exactly zero and the
actual flavor mix is more like $\eta:1:0$ where $\eta$ may be a few (2
to 4)\%.

A third case is of sources which emit dominantly neutrons originating in
photo-dissociation of nuclei\cite{goldberg}.  Decay of neutrons leads to on initial pure
``$\beta$-beam'' of $\bar{\nu}_e$ with flavor mix of 1:0:0; again
contaminated by $\nu_\mu$ at a few \% level.

It is also of interest to consider the flavor content of the very highest
energy neutrinos, sometimes called ``the GZK neutrinos'', which were
predicted\cite{venya} soon after the original observation of the GZK cutoff
mechanism\cite{greisen}. These are the
neutrinos emitted following the scattering of highest energy cosmic rays
on the microwave background photons. The dominant process is the production
of the resonance $\Delta^{+}$ which decays into a neutron and a $\pi^{+}$.
Below about 100 PeV, the neutrinos from neutron decay dominate, resulting
in the flavor mix $1: 0 :0$ whereas  above 100 PeV, pion decays dominate,
resulting in the conventional flavor mix of $2: 1: 0$ \cite{engel}.

\section{Effect of Oscillations}

The current knowledge of neutrino masses and mixing can
be summarized
as follows\cite{concha}. The mixing matrix elements are
given to a very good approximation by the so-called tri-bi-maximal
matrix\cite{harrison}.
The bound on the element  $|U_{e3}|$ comes from the CHOOZ experiment\cite
{chooz} and is given by $|U_{e3}| < 0.17 $.
The mass spectrum has two possibilities: normal or 
inverted. 
The mass differences are given by $|\delta m^2_{32}| \sim 
2.4.10^{-3} eV^2$(with the + sign corresponding to normal
hierarchy and - sign to the inverted one) and 
$\delta m_{21}^2 \sim +7.6.10^{-5} eV^2$.  Since $\delta m^2 
L/4E$ for 
the distances to GRB's and AGN's (even for energies up to 
and beyond PeV) 
is very large $(> 10^7)$ the oscillations have always 
averaged out 
and the conversion(or survival) probability is given by
\begin{eqnarray}
P_{\alpha \beta} &=& \sum_{i} | U_{\alpha i} \mid^2 \mid 
U_{\beta i} \mid^2 \
\end{eqnarray}
Assuming no significant matter effects en-route, it is easy 
to show that
the tri-bi-maximal mixing matrix  leads to a simple propagation matrix 
P, which, for any value of the solar mixing angle, 
converts a 
flux ratio of 
$\nu_e: \nu_\mu: \nu_\tau = 1:2:0$ 
into one of $1:1:1$.  Hence the flavor mix expected at 
arrival 
is simply an equal mixture of $\nu_e, \nu_\mu$ and 
$\nu_\tau$ as was observed long ago\cite{learned1,athar}.
If this universal flavor mix is confirmed by future 
observations, our current
knowledge of neutrino masses and mixing is reinforced and 
conventional
wisdom about the beam dump nature of the production 
process is confirmed as
well. However, it would be much more exciting to find 
deviations from it, and learn something new. How can this 
come about? I give below  a shopping list of  variety of ways 
in which this could come to pass, and what can be learned in each case.
  
\section{Deviations from Canonical Flavor Mix}

There are quite a few ways in which the flavor mix can be 
changed from
the simple universal mix. 

The first and simplest is that initial flavor mix is NOT
$1:2:0$. The damped muon case in which the initial flavor mix is
$0:1:0$,  the final result after the averaged out oscillations becomes
$0.57:1:1$ on arrival. The ``beta''  beam which starts out as
$1:0:0$ initially becomes $2.5:1:1$ on arrival. These are sufficiently
different from the universal mix so that  the nature of the
source can be easily distinguished from such observations. The
two kinds of production processes which both lead to the initial
flavor mix of $1: 2: 0$, namely the pp and $\gamma$p can also be distinguished
from each other, at least in principle\cite{goldberg1}.
In the former case the flux of $\bar{\nu}_e$ relative to the
total neutrino flux is given by 1/6, whereas in the latter case it is given 
by 2/27; the $\bar{\nu}_e$ flux can be measured at an 
incident energy of 6.3 PeV by the showers due to the ``Glashow''
resonance as was first stressed in Ref.7.

Small deviations from 1:1:1 can be used to determine
deviations from the tri-bi-maximal neutrino mixing, e.g.
a non-zero value for $U_{e3}$\cite{xing}. A non-zero
$U_{e3}$ leads to a flavor mix of
$(1+2\Delta):(1-\Delta):(1-\Delta)$ where $\Delta = \sqrt{2}/3
U_{e3} \cos(\delta)$. Also, for a pure
$\nu_{\bar{e}}$ beam,
the ratio R = $\mu/(e + \tau)$ is given approximately by
2/7 - 0.14$\Delta$, and hence is also a measure
of $|U_{e3}|$\cite{serpico}. Here it is assumed that $\theta_{23}$
is exactly $\pi$/4; one can also attempt precision measurement of
$\theta_{23}$\cite{serpico1}. In principle, one can also
envisage a damped muon beam to measure $\Delta$, as the ratio
R becomes  7/11  + 0.42$\Delta$. If the element $|U_{e3}|$ is known
by the time this measurement is made, one can hope to get a handle on the
CPV phase $\delta$\cite{nir}. Unfortunately, these measurements not only
need to measure small deviations but are
made even more  difficult by the impure nature of the initial flavor mixes as
has been discussed  recently\cite{lipari}. Another way in which
small deviations can arise is from small mixing with sterile neutrino
states\cite{choubey}.

The possibility that the mass differences between neutrino 
mass  eigenstates are zero in vacuum (and become non-zero 
only in the presence
of matter) has been raised\cite{kaplan}. If this 
is true, then the final flavor mix 
should be the same as initial, namely: $1:2:0.$ However,
very recently, analysis of low energy atmospheric neutrino
data by Super-Kamiokande has ruled out a wide variety of
models for such behavior\cite{SK}.

Neutrino decay is another important possible way for the 
flavor mix to 
deviate significantly from the democratic 
mix\cite{beacom}.  We now know that neutrinos have non-zero 
masses and non-trivial mixing,
based on the evidence for neutrino mixing and 
oscillations from the data on
atmospheric, solar and reactor neutrinos.

Once neutrinos have masses and mixing, then in general, the heavier neutrinos 
are expected to 
decay into the lighter ones via flavor changing 
processes\cite{pakvasa2}.  
The only remaining questions are (a) whether the lifetimes are 
short enough to be phenomenologically interesting (or are 
they too long?) and (b) what are the dominant decay modes.
Since we are interested
in decay modes which are likely to have rates (or lead to 
lifetimes)  which
are phenomenologically interesting,  we can rule out 
several classes of decay
modes immediately. For example, the very strong 
constraints on radiative decay
modes and on three body modes such as $\nu \rightarrow 
3\nu$ render them uninteresting.

The only decay modes 
which can have interestingly fast decay rates are two 
body 
modes such as $\nu_i \rightarrow \nu_j + x$  where 
$x$ is a very light or massless particle, e.g. a Majoron.
In general, the Majoron is a mixture of the Gelmini-
Roncadelli\cite{gelmini} and Chikasige-Mohapatra-
Peccei\cite{chicasige} type Majorons.  The effective 
interaction is of the form:
\begin{equation}
g \bar{\nu}^c_\beta (a+b \gamma_5) \nu_\alpha \ x
\end{equation}
giving rise to decay:
\begin{equation}
\nu_\alpha \rightarrow \bar{\nu}_\beta \ ( or  \ 
\nu_\beta)  +  x
\end{equation}
where  $\nu_\alpha$
and $\nu_\beta$
are mass eigenstates which may be mixtures of flavor and
sterile neutrinos.
Explicit models of this kind which can give rise to fast
neutrino decays have
been discussed\cite{valle}.
The models with $\Delta L = 2$ are unconstrained by $\mu$ and $\tau$ decays
which cannot be engendered by such couplings.
Both($\Delta L = 2$ and $\Delta L = 0$) kinds of models with couplings  
of $\nu_\mu$ and $\nu_e$ are constrained
by the limits on multi-body $\pi$, K decays,
and on $\mu-e$ universality violation in $\pi$ and K 
decays\cite{barger}, 
but these bounds allow fast neutrino decays.

There are a number of  interesting cosmological implications of
such couplings. The
details depend on the mass spectrum of neutrinos and the
scalars in the model, and on the strength of the couplings.
For example, when all the neutrinos are heavier than the
scalar; for sufficiently strong coupling(g$ > 10^{-5}$) the relic
neutrino density vanishes today, and the neutrino mass 
bounds from CMB
and large scale structure are no longer operative, in the sense 
that potentially strong cosmological bounds can be violated by future
measurements in the laboratory which  find a non-zero 
result for a neutrino mass\cite{beacom1}. 
If the scalars are heavier than the neutrinos, there are 
signatures
such as shifts of the $n$th multi-pole peak (for large $n)$ 
in the 
CMB\cite{chacko}.
There are other implications as well, such as the number 
of relativistic
degrees of freedom(or effective number of neutrinos) being 
different at the
BBN and the CMB eras. The additional degrees of freedom 
should be detectable
in future CMB measurements. The CMB data show preferance for some free 
streaming neutrino components(but not all species) to be present during
the photon decoupling era.
If this could be established for all three flavors, very stringent 
limits on the couplings and hence on neutrino lifetimes can
be derived, although this is not possible at present\cite{bell}.

Direct limits on such decay modes are rather weak.
Current bounds on such decay modes are as follows.  For 
the mass eigenstate $\nu_1$, the limit is about
\begin{equation}
\tau_1 \geq 10^5 \ sec /eV
\end{equation}
based on observation of $\bar{\nu}_e's$ from SN1987A 
\cite{hirata}
(assuming CPT invariance). For $\nu_2$, strong  limits can 
be deduced  from
the non-observation of solar anti-neutrinos in 
KamLAND\cite{eguchi}.
A more general but similar bound is obtained from 
an analysis of solar neutrino data\cite{bell1}. This 
bound is given by:
\begin{equation}
\tau_2 \geq 10^{-4} \ sec/eV
\end{equation}
For $\nu_3$,  one can derive 
a bound from the atmospheric neutrino observations of 
upcoming neutrinos\cite{barger1}:
\begin{equation}
\tau_3 \geq \ 10^{-10} \ sec/eV
\end{equation}

The strongest lifetime limit is thus too weak to eliminate 
the possibility of
astrophysical neutrino decay by a factor about $10^7 
\times (L/100$ Mpc) 
$\times (10$ TeV/E).  It was noted that the
disappearance of all states except $\nu_1$ would prepare a 
beam that could in principle be used to measure elements 
of the neutrino mixing matrix\cite{pakvasa3}, namely the ratios $|U_{e1}|^2 
: |U_{\mu 1}|^2 : |U_{\tau 1}|^2$.  
The possibility of measuring
neutrino lifetimes over long baselines was mentioned in 
Ref.\cite{weiler}, 
and some predictions for decay in four-neutrino models 
were given in 
Ref.\cite{keranen}.  The particular values and small 
uncertainties on the neutrino mixing parameters allow 
for the first time very distinctive signatures of the 
effects of 
neutrino decay on the detected flavor ratios.  
The expected increase in neutrino lifetime sensitivity 
(and corresponding 
anomalous 
neutrino couplings) by several orders of magnitude makes 
for a very
interesting test of physics beyond the Standard Model; a 
discovery would
mean physics much more exotic than neutrino mass and 
mixing alone.   
Because of its unique signature, neutrino decay  cannot be 
mimicked by either different neutrino flavor ratios at the 
source or other non-standard neutrino interactions.

A characteristic feature of decay is its strong energy 
dependence: 
$\exp (-Lm/E \tau)$, where $\tau$ is the rest-frame 
lifetime.  
For simplicity, we will consider the case that decays are always 
complete, i.e., that 
these exponential factors vanish.  
The simplest case (and the most generic expectation) is a 
normal hierarchy 
in which both $\nu_3$ and $\nu_2$ decay, leaving only the 
lightest stable eigenstate  $\nu_1$.  In this case the 
flavor ratio is\cite{pakvasa3} 
$|U_{e1}|^2:  |U_{\mu 1}|^2 : |U_{\tau 1}|^2$. 
Thus, if $|U_{e3}| = 0$ we have
\begin{equation}
\phi_{\nu e} :  \phi_{\nu_{\mu}} :  \phi_{\nu_{\tau}}
\simeq 4 : 1 : 1, 
\end{equation}
where we used the propagation matrix derived from the tri-bi-maximal mixing.  
Note that this is an extreme deviation of the flavor ratio 
from
that in the absence of decays.  It is difficult to imagine 
other mechanisms
that would lead to such a high ratio of $\nu_e$ to 
$\nu_\mu$.  In the case
of inverted hierarchy, $\nu_3$ is the lightest and hence 
stable state, and
so\cite{beacom} we have instead 
\begin{equation}
\phi_{\nu_{e}} :  \phi_{\nu_{\mu}} : \phi_{\nu _{\tau}} = 
|U_{e3}\mid ^2 : 
|U_{\mu 3} \mid^2 : |U_{\tau 3} \mid^2 = 0 : 1 : 1.
\end{equation}
If  $|U_{e3}| = 0$ and $\theta_{atm} = 45^0$, each mass 
eigenstate has equal
$\nu_\mu$ and $\nu_\tau$ components.  Therefore, decay 
cannot break 
the equality between the $\phi_{\nu_{\mu}}$ and 
$\phi_{\nu_{\tau}}$ 
fluxes and thus the $\phi_{\nu_{e}} : \phi_{\nu_\mu}$ 
ratio contains all the useful information.

When $|U_{e3}|$ is not zero, and the hierarchy is normal, it 
is possible to
obtain information on the values of $|U_{e3}|$ as well as 
the CPV phase $\delta$\cite{beacom2}.  The flavor ratio 
$e/\mu$ varies from  4 to 10 (as $|U_{e3}|$ goes from 0 to 
0.2)  
for $\cos \delta =+1$ but from 4 to 2.5 for $\cos \delta =-
1$.  The ratio $\tau/\mu$ varies from 1 to 4 $(\cos \delta 
= +1)$ or 1 to 0.25 $(\cos \delta =-1)$ for the same range 
of $U_{e3}$.

If the decays are not complete and if the daughter does 
not carry the full
energy of the parent neutrino; the resulting flavor mix is 
somewhat
different but in any case it is still quite distinct from the 
simple $1:1:1$
mix\cite{beacom}. There is a very recent exhaustive study
of the various possibilities\cite{winter}.

If the path of neutrinos takes them thru regions with 
significant magnetic 
fields and the neutrino magnetic moments are large enough, 
the flavor mix can 
be affected\cite{enquist}.  The main effect of the passage 
thru magnetic field is the 
conversion of a given helicity into an equal mixture of 
both helicity states.
This is also true in passage thru random magnetic 
fields\cite{domokos}. It has been shown recently that the presence of
a magnetic field of a few(10 or mor) Gauss at the source can make
the neutrinos decohere as they traverse cosmic distances\cite{farzan}.

If the neutrinos are Dirac particles, and all magnetic 
moments are comparable, 
then the effect of the spin-flip is to simply reduce the 
overall flux of all 
flavors by half, the other half becoming the sterile Dirac 
partners.
If the neutrinos are Majorana particles, 
the flavor composition remains 1 : 1 : 1 when it 
starts from 1 : 1 : 1, and the absolute flux remains 
unchanged.

What happens when large magnetic fields are present in or 
near the neutrino 
production region?  In case of Dirac neutrinos, there is 
no difference and 
the outgoing flavor ratio remains 1 : 1 : 1, with the 
absolute fluxes
reduced by half.  In case of Majorana neutrinos,
since the initial flavor mix is no longer universal but is 
$\nu_e: \nu_\mu: \nu_\tau \approx 1: 2: 0,$ this is 
modified
but it turns out that the final(post-oscillation) flavor 
mix is still 1 : 1
: 1!

Other neutrino properties can also affect the neutrino 
flavor mix and modify
it from the canonical 1 : 1 : 1. If neutrinos have 
flavor(and equivalence
principle) violating couplings to gravity(FVG); then there can be resonance
effects which make for one way transitions(analogues of 
MSW transitions)
e.g. $\nu_\mu \rightarrow \nu_\tau$ but not vice
versa\cite{minakata,barger3}. In case of FVG for example,
this can give rise to an anisotropic deviation of the 
$\nu_\mu/\nu_\tau$
ratio from 1, becoming less than 1 for events coming from 
the direction
towards the Great
Attractor, while remaining 1 in other 
directions\cite{minakata}. If such striking effects are not seen,
then the current bounds on such violations can be improved by
six to seven orders of magnitude.
 
Another  possibility
that can give rise to deviations of the flavor mix from 
the canonical
1 : 1 : 1 is the idea of neutrinos of varying 
mass(MaVaNs). In this
proposal\cite{fardon}, by having the dark energy and 
neutrinos(a sterile one to be
specific) couple, and track each other; it is possible to 
relate the
small scale ($2\times 10^{-3}$ eV) required for the dark 
energy to the small
neutrino mass, and furthermore the neutrino mass depends 
inversely
on neutrino density, and hence on the epoch. As a result, 
if this
sterile neutrino mixes with a flavor neutrino, the mass 
difference
varies along the path, with potential resonance 
enhancement of the
transition probability into the sterile neutrino, and thus 
change the
flavor mix\cite{hung}. For example, if only one 
resonance is crossed en-route, it can 
lead to a conversion of the heaviest (mostly) 
flavor state into the (mostly) sterile
state, thus changing the flavor mix to
 $1-|U_{e1}| ^2 \ : 1-|U_{\mu 1}|^2 \ : 
1-|U_{\tau 1}|^2
\approx 0.4 \ : 1 \ : 1,$ in case of inverted
hierarchy and to 
$1-|U_{e3}| ^2\ : 1-|U_{\mu 3}| ^2 \: 1-|U_{\tau3}|^2 \approx 2 \ : 1 \ : 1$
in case of normal hierarchy.

Complete quantum decoherence would give rise to a flavor 
mix given
by $1:1:1$, which is identical to the case of averaged out 
oscillations
as we saw above. The distinction is that complete 
decoherence always
leads to this result; whereas averaged out oscillations 
lead to this
result only in the special case of the initial flavor mix 
being $1:2:0.$
To find evidence for decoherence, therefore, requires a 
source which
has a different flavor mix . One possible practical 
example is the ``beta'' beam  source
with an initial flavor mix of $1:0:0$. In this 
case  decoherence 
leads to the universal $1:1:1$ mix whereas the averaged 
out oscillations
lead to $2.5:1:1$\cite{hooper}. The two cases can be easily 
distinguished from each other.

Violations of Lorentz invariance and/or CPT invariance
can change the final flavor mix from the canonical
universal mix of $1: 1: 1$ significantly. With a specific
choice of the change in dispersion relation due to Lorentz
Invariance Violation, the effects can be dramatic. For example,
the final flavor mix at sufficiently high energies can become
$ 7: 2: 0$\cite{hooper}.

If each of the three neutrino mass eigenstates is actually 
a doublet 
with very small mass difference (smaller than $10^{-6} 
eV)$, 
then there are no current experiments  that could have 
detected this. 
Such a possibility was raised long ago\cite{bilenky1}. 
It turns out that the only way to detect such small mass 
differences $(10^{-12} eV^2 > \delta m^2 > 10^{-18} eV^2)$ 
is by measuring flavor mixes of the high energy neutrinos 
from cosmic sources.  Relic supernova neutrino
signals and AGN neutrinos are sensitive to mass difference 
squared down to $10^{-20} eV^2$ \cite{beacom3}.

Let $(\nu_1^+, \nu_2^+, \nu_3^+; \nu_1^-,  \nu_2^-, 
\nu_3^-)$ 
denote the six mass eigenstates where $\nu^+$ and $\nu^-$ 
are a 
nearly degenerate pair.  A 6x6 mixing matrix rotates the 
mass 
basis into the flavor basis.  
In general, for six Majorana neutrinos, there would be 
fifteen 
rotation angles and fifteen phases.  However, for pseudo-
Dirac 
neutrinos, Kobayashi and Lim\cite{kobayashi} have given an 
elegant proof 
that the 6x6 matrix $V_{KL}$ takes the very simple form 
(to lowest order in $\delta m^2 / m^2$:
\begin{eqnarray}
V_{KL} = \left (
\begin{array}{cc}
U & 0 \\
0 & U_R 
\end{array} \right) \cdot
\left (
\begin{array}{cc}
V_1 & iV_1 \\
V_2  & -iV_2 
\end{array}\right),
\end{eqnarray}
where the $3\times 3$ matrix U is just the usual mixing 
(MNSP)matrix determined by the atmospheric and solar 
observations, the 
$3\times 3$ matrix $U_R$ is an unknown unitary matrix and 
$V_1$ and 
$V_2$ are the diagonal matrices 
$V_1 =$ diag $(1,1,1)/\sqrt{2}$, and $V_2$=diag$(e^{-i 
\phi_1}, 
e^{-i \phi_2}, e^{-i \phi_3})/\sqrt{2}$, with the $\phi_i$ 
being arbitrary phases.

As a result, the three active neutrino states are 
described in terms of the six mass eigenstates as:
\begin{equation}
\nu_{\alpha L} = U_{\alpha j} \ \frac{1}{\sqrt{2}} 
\left(\nu^+_{j} + i \nu^-_{j}  \right).
\end{equation}

The flavors deviate from the democratic value of 
$\frac{1}{3}$ by
\begin{eqnarray*}
\label{deviate3}
\delta P_e &=& -\frac{1}{3}\,\left[ \frac{3}{4}\chi_1 + 
\frac{3}{4}\,
\chi_2 \right],\nonumber\\
\delta P_\mu = \delta P_\tau  &=& -\frac{1}{3}\,\left[ 
\frac{1}{8}\chi_1 + \frac{3}{8}\,\chi_2 
	+ \frac{1}{2}\,\chi_3\right] \,
\end{eqnarray*} 
where $\chi_i = \sin^2(\delta m_i^2 L/4E)$.The flavor 
ratios deviate 
from $1:1:1$ when
one or two of the pseudo-Dirac oscillation modes is 
accessible.  In
the ultimate limit where $L/E$ is so large that all three 
oscillating
factors have averaged to $\frac{1}{2}$, the flavor ratios 
return to $1:1:1$,
with only a net suppression of the measurable flux, by a 
factor of
$1/2$. As a bonus, if such small pseudo-Dirac mass differences 
exist, it would enable us to measure cosmological parameters such as
the red shift in neutrinos(rather than in photons)\cite{weiler,beacom3}.

\section{Experimental Flavor Identification} 

It is obvious from the above discussion that flavor 
identification is
crucial for the purpose at hand. In a water(or ice) cerenkov 
detector flavors can be 
identified as follows.  

The $\nu_\mu$ flux can be measured by the $\mu's$ produced 
by the charged 
current interactions and the resulting $\mu$ tracks in the 
detector which
are long  at these energies.  $\nu_{e}'{s}$ produce 
showers by both
CC and NC interactions.  The total rate for showers 
includes those
produced by NC interactions of $\nu_\mu's$ and 
$\nu_\tau's$ as well and those
have to be (and can be) subtracted off to get the real 
flux of $\nu_e's$.
Double-bang and lollipop events are signatures unique to 
tau neutrinos, made
possible by the fact that tau leptons decay before they 
lose a significant
fraction of their energy.  A double-bang event consists of 
a hadronic shower
initiated by a charged-current interaction of the 
$\nu_\tau$ followed by a
second energetic shower from the decay  of the
resulting tau lepton\cite{learned1}.  A lollipop event 
consists of the second
of
the double-bang showers along with the reconstructed tau 
lepton track (the
first bang may be detected or not).  In principle, with a 
sufficient number
of
events, a fairly good estimate of the flavor ratio $\nu_e: 
\nu_\mu: \nu_\tau$ 
can be reconstructed, as has been discussed recently.
Deviations of the flavor ratios 
from $1:1:1$ due to possible decays are so extreme that 
they should be
readily identifiable\cite{beacom4}. Future high energy 
neutrino telescopes,
such as Icecube\cite{karle}, will not have perfect ability 
to separately 
measure the neutrino flux in each flavor.  However, the 
situation is
salvageable. In the limit of $\nu_\mu - \nu_\tau$ symmetry 
the fluxes for $\nu_\mu$ and 
$\nu_\tau$ are always in the ratio 1 : 1, with or without 
decay. 
This is useful since the $\nu_\tau$ flux is the hardest to 
measure. 

Even when the  tau events are not all
identifiable, the relative number of shower events to
track events can
be related to the most interesting quantity for testing
decay scenarios,
i.e., the $\nu_e$ to $\nu_\mu$ ratio.  The precision of
the upcoming
experiments should be good enough to test the extreme
flavor ratios produced
by decays.  If electromagnetic and hadronic  showers can
be separated, then
the precision will be even better\cite{beacom4}.Comparing,
for example, the
standard flavor ratios of 1 : 1 : 1 to the
possible 4 : 1 : 1 (or $ 0 : 1: 1$ for inverted hierarchy)generated by
decay, the higher(lower)
electron neutrino
flux will result in a substantial increase(decrease) in the relative 
number of 
shower events.The measurement will be limited only by the 
energy resolution of the
detector and the ability to reduce the atmospheric 
neutrino background(which 
drops rapidly with energy and 
should be negligibly small at and above the PeV scale).

\section{Discussion and Conclusions}  

The flux ratios we discuss are energy-independent to the extent 
that the following assumptions are valid: (a)the ratios at production are energy-
independent, (b) all oscillations are averaged out, and (c) 
that all possible decays are complete.  In the standard 
scenario with only oscillations, the final flux ratios are 
$\phi_{\nu_{e}} :  \phi_{\nu_{\mu}} : 
 \phi_{\nu_{\tau}} = 1 : 1 : 1$.  In the cases with decay, 
we have found rather
different possible flux ratios, for example 4 : 1 : 1 in 
the normal hierarchy and 
0 : 1 : 1 in the inverted hierarchy.  These deviations 
from 1 : 1 : 1 
are so extreme that they should be readily measurable.

If we are very fortunate, we may be able 
to observe a
reasonable number of events from several sources (of known 
distance) 
and/or over a sufficient range in energy.  Then the 
resulting 
dependence of the flux ratio $(\nu_e/\nu_\mu)$ on L/E as 
it evolves from say 4 (or 0) to 1 can 
be clear evidence of decay and further can pin down the 
actual lifetime
instead of just placing a bound\cite{barenboin}.

To summarize, we suggest that if future measurements of 
the flavor mix at
earth of high energy astrophysical neutrinos find it to be
\begin{equation}
\phi_{\nu_{e}} / \phi_{\nu_{\mu}} / \phi_{\nu_{\tau}} = 
\alpha / 1 / 1 ;
\end{equation}
then
\begin{itemize}
\item[(i)] $\alpha \approx 1$ (the most boring case) 
confirms our knowledge of the
MNSP\cite{MNS} matrix and our prejudice about the 
production mechanism;
\item[(ii)] $\alpha \approx 1/2$ indicates that the source 
emits pure
$\nu_\mu's$ and the mixing is conventional;
\item[(iii)]$\alpha \approx 3$ from a unique direction, 
e.g. the Cygnus region, would be
evidence in favor of a pure $\bar{\nu}_e$ production as 
has been suggested
recently\cite{goldberg};
\item[(iv)] $\alpha > 1$ indicates that neutrinos are 
decaying with normal
hierarchy; and 
\item[(v)]$\alpha \ll 1$ would mean that neutrino decays 
are occurring with
inverted hierarchy;
\item[(vi)] Values of $\alpha$ which cover a broader range 
(2.5 to 10) and 
deviation of the $\mu/\tau$ ratio from 1(between 0.2 to 4) 
can yield valuable 
information about $U_{e3}$ and $\cos \delta$. Deviations 
of $\alpha$
which are less extreme(between 0.7 and 1.5) can also probe 
very small pseudo-Dirac 
$\delta m^2$ (smaller than $10^{-12} eV^2$).
\end{itemize}  

Incidentally, in the last three cases, the results have 
absolutely no
dependence on the initial flavor mix, and so are 
completely free of any
dependence on the production model. So either one learns 
about the production
mechanism and the initial flavor mix, as in the first 
three cases, or one
learns only about the neutrino properties, as in the last 
three cases.
To summarize, the measurement of neutrino flavor mix at 
neutrino telescopes
is absolutely essential to uncover new and interesting 
physics of neutrinos. 
In any case, it should be evident that the construction of 
very large neutrino telescopes  is a ``no lose'' 
proposition.

\section{Acknowledgments}
This talk is based on work done in collaboration with
John Beacom, Nicole Bell, Dan Hooper, John Learned, Werner
Rodejohann and 
Tom Weiler. I thank them for a most
enjoyable collaboration and Tom Weiler for a very careful reading of 
the manuscript. I would like to thank  the 
organizers of COSPA 2007 for the
opportunity  to present this talk as well as their hospitality 
and for providing 
a most stimulating atmosphere during the meeting.
This work was supported in part by U.S.D.O.E. under grant 
DE-FG02-04ER41291.

\end{document}